# Role of Management Information System in Time Saving: A Case Study of Automobile Tax System in Sindh.

Muhammad Bilal Liaquat,
Department of Computer Sciences,
Bahria University Karachi Campus,
Karachi, Pakistan.

*Abstract*: This Case Study will be used in order to investigate and establish the importance of role of Management Information System in time saving during the payment of automobile tax in Sindh through e-filling methods. Moreover it will also highlight the important factors which are involved as barriers and limit the role of MIS in time saving techniques. The approach which is used in this case study is descriptive research type along with the survey. The data used was collected from the specimen of common people working in different environments along with the officers working at Civic Centre (Automobile Tax Collection Branch *excise department*). The audience included were all well informed by the process and were eligible to give their opinions on the following research. A system design is also proposed along with an Erd which can be useful in the coming future. This research could likewise be expanded to include different of respondents, for example, paid taxpayers and different types of taxpayers. Paid tax payers are given the rights by their clients to prepare their assessment matters. They use the e-filing system for different types of clients and are more frequent users of the e-filing system than taxpayers who file for themselves. It would be interesting to understand which facets of hazard are larger to them. Different types of taxpayers, for example, company authorized cars may deal with more complex exchanges than single car taxpayers, consequently, they may emphasize different hazard facets when filing in the government form frame electronically.

## I. Introduction

Governments around the world are increasing the use of data and communication technologies to improve the delivery of public services and the distribution of public organization data to people in general. Manual tax filling remains the traditional and most widespread method of submitting individual automobile tax for government revenue services, in Pakistan or any other country. For quite a while, both the United States and Canada have attempted to acquaint electronic filing frameworks with enhance government operations and decrease costs [1]. Regardless of the considerable number of endeavors went for the advancement of better and simpler electronic assessment filing frameworks, these expense filing frameworks stayed unnoticed by the general population or were genuinely underused disregarding their accessibility. Therefore, there is a need to comprehend the acknowledgment by the clients of the electronic duty filing frameworks and distinguish the components that can influence their choice to utilize or not utilize these electronic assessment filing frameworks. This issue is imperative in that the appropriate response could help the administration to design and advance new types of electronic assessment filing frameworks later on. In an era where Information Technology is growing rapidly, the right usage of that technology has become a primary goal for every organization. The expanding enthusiasm for MIS has prompted much action in creating methods and programming for information administration [2]. Late research has demonstrated that "trust" has a striking influence on client eagerness to take part in online trades of cash and individual delicate data. Besides, e-documenting offers many advantages to specialist organizations, which are the assessment experts. The specialist co-op, e-recording limits their workload and operational cost because of the settlement of cost frames in a paperless situation. It additionally reduces the cost of processing, putting away and treatment of assessment forms. Despite these benefits associated with e-filing, charge authorities face some real challenges towards the implementation of the e-filing system Along these lines, perceived ease of use and perceived usefulness may not appropriately disclose the user's goal to embrace the electronic duty filing framework. Thus it ends up plainly important to look for extra factors that can better foresee the acknowledgment of electronic duty filing frameworks. It is due to the techniques involved in MIS that everyone using its features are saving an ample amount of time. Recently, Malaysia also introduced tax e-filing (also referred to as e-filing in this paper). In 2009, the fourth year after e-filing was implemented, only 1.25 million taxpayers were reported to have filed their tax return through e-filing (Bernama, 2009) [3]. Researches such as Hoffman et al. (1995), Alba et al. (1997) and Peterson et al. (1997) have discussed several benefits of online activities to the consumers. Among them are that the Internet allows consumers to conduct transactions within a few mouse clicks. E-filing also offers flexibility of time and reduces calculation error on the tax return form to the taxpayers. On the off chance that duty specialists are not ready to give an e-filing framework that could conquer these difficulties, citizens may be hesitant to receive the e-filing framework. In case duty specialists are not ready to give an e-filing framework that could conquer these difficulties, citizens might be hesitant to receive the e-filing framework. The issues, for example, loss of imperative time, information protection, glitches on the framework's execution, if not deliberately overcome, could be made an interpretation of into dangers to present and potential adopters of the e-filling framework. Along these lines, danger of e-filling is characterized as the general measure of vulnerability or nervousness observed by a person in using e-filling. It is essential for the authorities to understand the dangers noticed by the citizens and to guarantee that these dangers are limited for the effective execution of the e-filling framework. After using an e-service over the Internet, the public may find the e-service system easy and useful or otherwise. Since the public cannot directly communicate with tax personnel, see or touch the tax forms as the service is provided online, the e-filing service system delivered to them may not perform as expected. In addition, the public may be burdened by the time and effort spent learning the new system and accommodating any services failure. Although time is a non-monetary effort and varies among individuals, researchers have recognized that time is a cost that consumers/users must pay for any use of products/services [4] (Sweeney et al., 1999). Based on the effective evaluation of goal-centered information systems, the following paper is divided into following different parts. Firstly we evaluate the effect of **Automobile tax payment system in Sindh,** which will help in the cause of time saving. Secondly we will also discuss the barriers involved in limiting the role of MIS. In this fasting moving and web oriented world it is very hard for anyone to visit a specific place in order to pay their taxes when we can easily pay for almost everything through a web-based application. With the help of some stats gathering we will examine the problem using **mean standard deviations**, **percentages**, **frequency counts**, test along with a prototype of a database erd which will in future help them to make an efficient system which can hold their data easily. By explaining the usage intention from the perspective of the users, the findings of this research not exclusively can help the government authorities to develop a better user-accepted electronic expense filing system, yet in addition provide bits of knowledge into how to promote the new IT to potential users [5].

## II. Background Study

Soon after partition, a ticket collector was making the rounds of a train and asked one of the passengers to show him his ticket. The passenger got angry and shouted, "Are you out of your mind? Haven't you heard that we are 'Free' now? Why should we purchase tickets anymore?" This story might be fictitious but it generally reflects the attitude of Pakistani nation towards payment of taxes [6]. Tax avoidance is an endemic disease in our part of the world. People would go to all lengths just to save a few bucks in taxes. On one hand, talk to the person on the street and they will decry the pathetic situation of public services like government hospitals, educational institutes and law enforcement agencies. Ask the same person about tax returns and you will receive lengthy rants about inefficient governance but not a peep about paying taxes. The fact that only 0.3% (500,000 people) of the population of Pakistan files their income tax returns does not raise many eyebrows because of our national attitude towards this issue. Taxation in Pakistan is a very complex system including more than 70 different taxes and nearly 37 different government agencies administer the tax system. Around 10 million people are registered to pay taxes but only 1.9 million people pay taxes. In 2002, Transparency International studied 256 respondents, among which 99% were facing corruption regarding to taxation. Furthermore, 32% of respondents were paying bribes to lower their tax assessment, and around 14% reported receiving fictitious tax assessments. Pakistan is one of the developing countries where tax to GDP ratio is very [7]. According to the stats, only 1.21 million citizens pay income tax making it less than 1% of the total population. But according to state bank of Pakistan's annual report, 57.5 million people are employed and obviously are earning some income and thus 57.5 million people should be paying income tax one way or the other. So, that rules out the figure of tax payers that keeps on floating here and there. Moreover, Pakistan's total population is around 200 million. According to the common notion, 29% of the population is paying the taxes. But according to Economic survey of Pakistan. 61.4% population is of working age making 122 million people fall into the working population. The 57.5 million tax payers constitute of 48% these working people [8]. Roughly 140 million portable clients pay over 20% tax on their utilization. Individuals utilize vehicles which utilize non-renewable energy sources that are liable to various expenses and therefore individuals wind up paying them. These are the sort of expenses that nobody can dodge. These expenses are without a doubt paid out of individual salaries. If we look into the old traditional way in which the government of Sindh is collecting automobile tax without keeping the customer comfort in mind. Our present tax collecting framework is extremely weak which contains escape clauses in the framework joined by the defilement of Sindh Board of Revenue (SBR) officers which talks about the equity and value of our administration authorities for their people. Subsequent to breaking down the tallness of debasement, uncommon support given to government authorities or their companions, our undocumented economy and inability to build assess net it has turned out to be important for any moneylender to set the objective keeping in mind the end goal to proceed with the cash loaning and to ensure that their obtained cash can only be reimbursed by the borrower. If you are new to this routine of paying automobile tax than you have two choices to pay your tax.

1. **Civic Centre Excise Department.**
2. **National Bank of Pakistan** (specific Branches)**.** The registration of vehicles is being effected through computerized system and tax is also collected through 12 designated National Bank branches, linked with Civic Centre Karachi.

*Table a*

| S.No | Branch Name | Phone Number | Address | STATUS |
|---|---|---|---|---|
| | NATIONAL BANK OF PAKISTAN KARACHI BRANCHES | | | |
| 1 | Awami Markaz | 021-99240558/ 99240556 | SHAHRAH-E-FAISAL KARACHI | A |
| 2 | Denso Hall Branch | 021-32620769-32620389 | Building Dense Hall M.A.Jinnah road | A |
| 3 | M.A. Jinnah Road | 021-99215025 - 99215026 | Kandawala Building M.A.Jinnah Road Karachi | A |
| 4 | Kehkashan Clifton | 021-99251330/99251344 | D-65, Block 9, KDA Scheme 5, Kehkashan Clifton | A |
| 5 | Defence Housing Society Branch | 021-35888259-35886334 | 15-A,Defence Housing Society Korangi Road Karachi | A |
| 6 | PIDC House | 021-99206031 to 33 | P IDC House Dr.Ziauddin Ahmed Road Karachi | A |
| 7 | Nazimabad Branch | 021-36623612 - 36607356 | 5-A,Ist Chowrangi Nazimabad | A |
| 8 | S.I.T.E. Karachi | 021-32567788 - 32573812 | Shopping Centre SITE Karachi | A |
| 9 | Korangi Industrial Area | 021-35062491 - 35071601 | Korangi SITE Karachi(Saudabad Merge Model Colony | A |
| 10 | Shaheed e milat | 021-34532771-34382569- | F/W.35-P/1,Karachi Co-operative Society Shaheed-e-Millat Road | A |
| 11 | Fatima Jinnah Road (Hub Branch) Hyderabad | 99200082-9200142-44 | NBP Building Fatima Jinnah Road Hyderabad | A |
| 12 | Gulshan Branch | 34521945 | *---Temporarily closed---* | NA |

And collect the Automobile tax from the car
Holders:                                    *Table b*

| S.No | Engine Capacity | Amount |
|------|-----------------|--------|
| 1 | Up to 1000 cc | Rs.750/- |
| 2 | 1001 cc to 1199 cc | Rs.1250/- |
| 3 | 1200 cc to 1299 cc | Rs.1750/- |
| 4 | 1300 cc to 1599 cc | Rs.3000/- |
| 5 | 1600 cc to 1999 cc | Rs.4000/- |
| 6 | 2000 cc and above | Rs.8000/- |

### III. Current Approach:

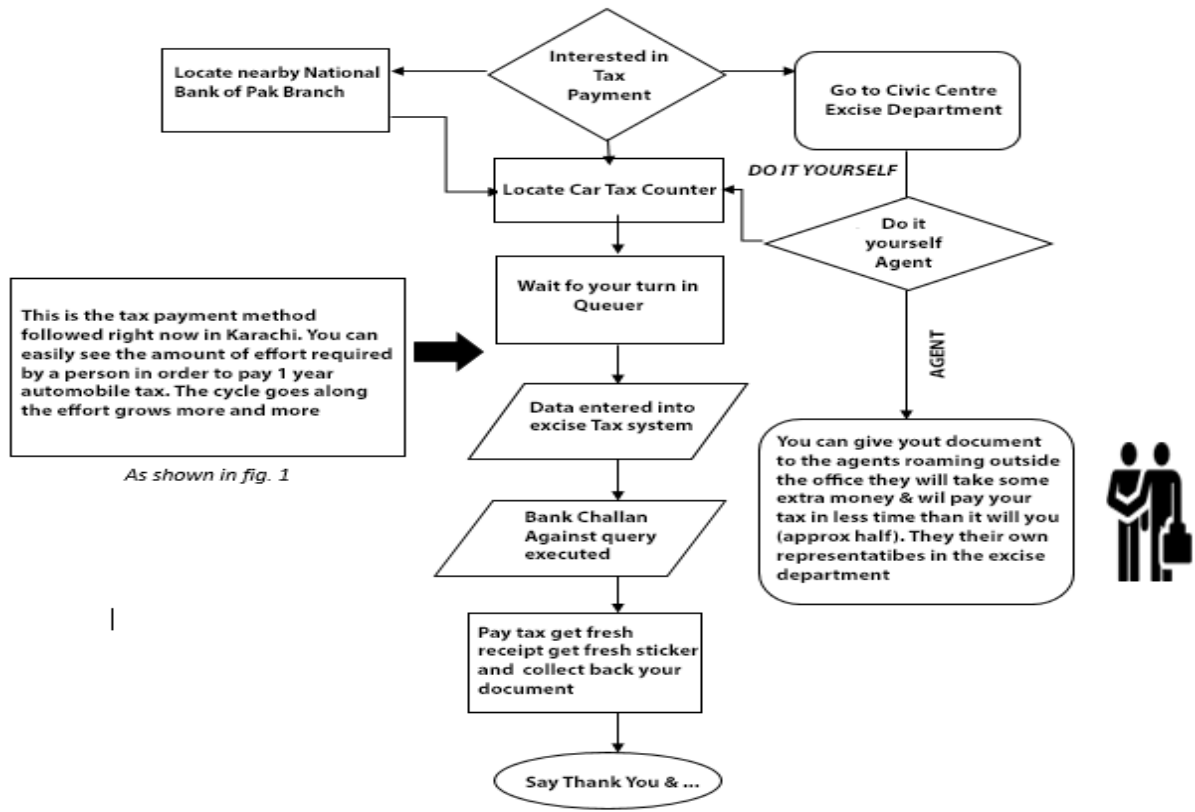

As shown in fig. 1

I personally visited both National Bank of Pakistan (*Awami Markaz branch*) and Civic Centre *excise department* in order to pay automobile tax and to gain personal experience so that I can discuss what happened there in detail. Firstly I went to National Bank of Pakistan (*Awami Markaz branch*) with ambitions of paying automobile tax. As I went inside the bank I humbly asked about the tax payment counter to a person who seemed to be an employee of the bank. He in quite an ignorant manner directed me towards a wall where I could easily see '*exit here*' written so I reached out towards another employee who finally directed me towards a counter where no one was sitting. I went to that counter and asked the guard about the officer and he replied to me that the officer will come after 10 a.m. We should keep one thing in mind that the official time for the officers to arrive to the bank is 8:30 a.m. but we know the Pakistani Government officials culture so I quietly sat down on the chair in order to wait for the official to arrive. He finally arrived and I reached out for him but he instructed me to wait for some more time as he had just arrived and wanted 10-15 minutes more to relax and settle in. He finally called me and I humbly went towards him and handed over my car's registration book which he took and entered some data into his computer and asked me that for how many years do I wish to pay my tax and told me the amount which he just got against my car details. He then handed over to me a challan and directed me towards a counter and told me to pay the required money there. I went there and as usual waited for the officer as he was busy drinking tea and having some refreshments. He finally took my challan along with the money and handed over to me my challan with 'amount paid' stamp on it. I went back towards the respected automobile tax collector and handed over to him the challan with the stamp on it. He in exchange gave me a sticker for 1 year, tax paid along with my registration book in which he marked a stamp as well and my part of the challan that I just paid. The following process took around **20 minutes** in which 25 minutes that I waited for the required person to arrive is not included. And the amount of time that I wasted to reach the nearest branch from home should also be kept in mind. If I roughly calculate the amount of time that was required in order to pay my automobile tax then keeping the traffic of Karachi in mind I approximately achieved my milestone in about 60-75 minutes. My second target was to visit Civic Centre *excise department* and calculate the amount of time required to complete my desired task. In this case I borrowed the original registration book of one of my friend and told him that I will pay his car tax for him for the sake of my information gathering process. He happily accepted my proposal and handed me over the registration book. As I parked my vehicle an agent came towards me and said that he has his contacts in the excise department and will pay my car tax and will take extra *300 rupees (Bribe or his service charges whatever you think)*. I first went into the department myself and was amazed to see a long queue of people waiting for their turn. I roughly estimated the amount of time required and it would have taken at least 1.5 hour for my turn to come so I came back downstairs and handed over my documents to the agent and he came after 15 minutes and handed over me my paid tax slip along with the sticker. This kind of mafia is usually normal in almost every government department and this practice is observed quite often and is totally normal.

**Purpose of the study:** The purpose of this study is to investigate and overcome such problems with the help of MIS in time saving techniques and decision making. It also highlights the constraints and obstacles involved in limiting the role of MIS in Government organizations.

## IV. Material & Methods involved:

The two fundamental develops that impact behavioral aim are Perceived Usefulness (PU) and Perceived Ease of Use (PEU). PU is characterized as the client's impression of how much utilizing the framework will enhance his or her execution in the work environment. PEU is characterized as the client's view of the measure of exertion they require, to utilize the framework. In particular, the paper proposes a model that examines the impact of PEU, PU and perceived risk of the tax eservice on the adoption behavior of taxpayers. Various e-service literature [9] indicates the significant effect of perceived risk on behavioral intention. Research such as Cunningham (1996), Bellman et al. (1999) and Featherman and Pavlou (2003) identified several facets of risk that are important in influencing behavior. Among these, Featherman and Pavlou's (2003) categorization of risks into its seven facets is found to be the most useful for evaluating the adoption of the e-filing system for various reasons [10]. The investigation utilized the descriptive research of the overview sort in type of surveys. A polls was planned and appropriated among individuals utilizing vehicle and are citizen also. The polls comprises of 3 sections. A section contained inquiries with respect to the individual data of the respondents, while part B contained 7 things on the utilization of MIS in Time sparing Techniques. The C part contained of 3 inquiries regarding the issues and hindrances that farthest point the part of MIS in efficient systems. The data gathered was coded and processed into a Statistical Software Package (SPSS). A total of 240 questionnaires were distributed and out of that 222 people returned the questionnaires that makes the response rate around 92.5 % (93)approx.), which was then forwarded for analysis. The result in the table below shows the analyzed result in tabular fo----

Table 1: Respondents by age

| Age | Frequency | Percent (%) |
|---|---|---|
| 18-24 | 14 | 6.3 |
| 25-34 | 46 | 20.7 |
| 35-44 | 72 | 32.4 |
| 45-49 | 42 | 18.9 |
| 50-59 | 38 | 17.1 |
| 60+ | 10 | 4.5 |

Table 2: Respondents by Qualification

| Qualification | Frequency | Percent (%) |
|---|---|---|
| Intermediate | 20 | 9.009 |
| Bachelor Degree | 110 | 49.5 |
| Master Degree | 80 | 36.3 |
| Ph.D. | 12 | 5.40 |

Table 2: Respondents by Qualification

| Qualification | Frequency | Percent (%) |
|---|---|---|
| Intermediate | 20 | 9.009 |
| Bachelor Degree | 110 | 49.5 |
| Master Degree | 80 | 36.3 |
| Ph.D. | 12 | 5.40 |

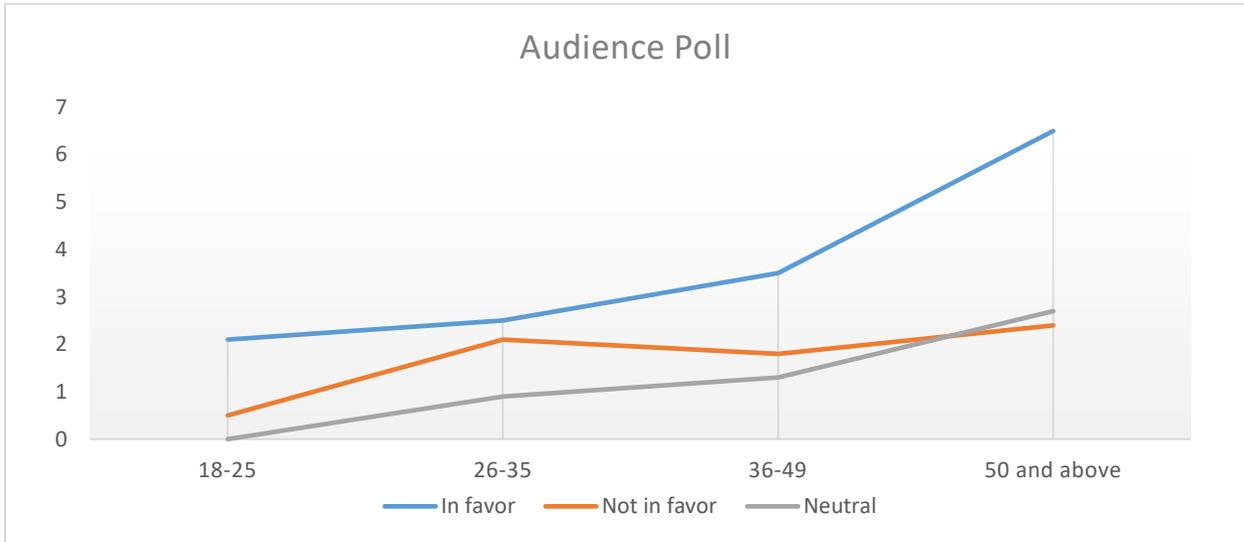

*Figure 2*

An audience poll was conducted and the results are *shown in figure 2*. We can clearly see that the people are in full favor of this change and are not happy by the current system.

## V. Proposed Framework:

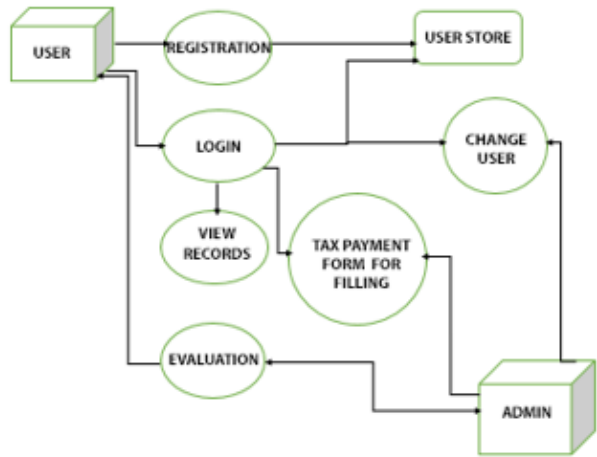

LEVEL 01 DATA FLOW DIAGRAM (DFD) FOR PURPOSED SYSTEM:

Figure 3)

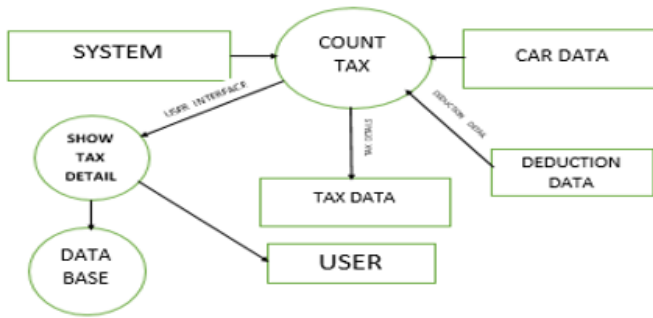

LEVEL 3 DATA FLOW DIAGRAM (DFD) FOR PURPOSED SYSTEM:

Figure 4

The proposed framework will consist of user panel *as shown in figure* 3, which will include login/registration facility for the user. Once you get login in a Tax payment portal will open and you will be asked to provide your car information and then the system will automatically calculate your automobile tax. Then you will be asked to provide your card information in order to pay your tax and once you have provided all the information required an e-receipt will be sent at your email. All your information will be secured through security firewall and different layers of network security. You can also see your old tax challans *as shown in figure 5* along with live support which will be monitored by the excise department support staff 24/7. Furthermore changes and advancement can also be made in the future. The frontend will be made in such a manner that a common person having just some knowledge about how to use a program can easily use it. The admin panel will be governed by the excise department and they will have the authority of making changes in user info, no third party will be given these rights. Both the admin and the user can evaluate their account. The user will add all their car's information and it will be saved in the database so that whenever the user logins in with his account in order to pay tax they should not worry about giving their car's information again and again.

The following is an **Entity Relationship Diagram** *(ERD)* which can be used in future in order to make an effective e-filling tax system for automobiles. This is just a prototype which can be helpful in making a system.

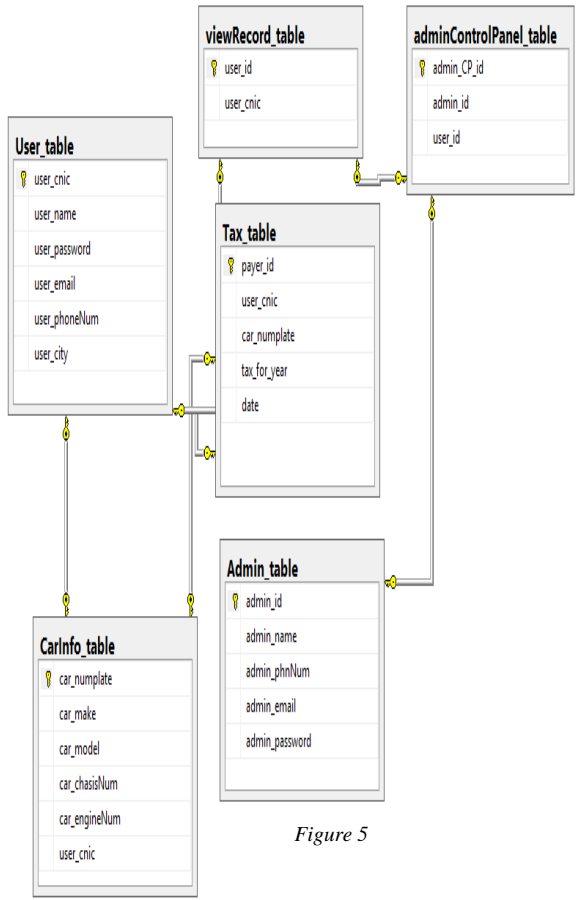

*Figure 5*

## VI. Functions of MIS involved in making of time saving techniques:

Given the fact that the usage of the electronic tax-filing systems is completely voluntary, and that the target user group consists of a large number of people with diversified backgrounds, the findings of this study suggest that in order to attract more users to use electronic tax-filing systems, it is not enough to develop a useful system and make the system easy to interact with. It is of paramount importance to develop electronic tax-filing systems that provide a solid trustworthy protection regarding security and privacy for the users. This way the government authorities need not concern themselves so much with attempting to directly influence behavioral intentions [10].To achieve the main objective the targeted audience were asked to identify the importance of MIS in time saving. It was clear that the audience were in full favor of importance of MIS in time saving.

    To achieve the first goal the audience were asked to rank up importance of MIS for time saving. Massive proportion of respondents (176 = 79.9%) believed that MIS provides a great backup for the organizations about what was done earlier for the betterment of time saving and what techniques were used before. This is because MIS contains information which is complete on the subject which allows the use of MIS in preparation of periodic reports.

In extension, (160 = 72.07 %) rated upon the fact that the information provided by the MIS is up-to-date and the system intercepts redundancy and repetition of data.

154 = 69.3% audience insisted upon the fact that after the introduction of MIS the performance of employees have also raised as they appear to be more efficient and are ready to compete with this technology. (190 = 85.5%) respondents think that the speed of extracting information from the system does not affect the accuracy of data extracted. The reliability of data should not questioned as there seems to be no doubt about the accuracy and efficiency of data extracted.

Furthermore, (147= 66.2%) ranked system provides correct information and is free of errors is easy to understand as it is arranged in order that any person how can read can understand the instructions and follow them easily.

In addition, (151 = 68.08) believe that further use of MIS in government departments will also help the *Audit* departments to make their reports accurate and reliable. While (173 = 78%) rated the fact that by using MIS in *excise department* we can make it corruption free as all the entries and data extracted and added will be done through systems and will make it easy for the higher authorities to fill in the loop holes if required in any scenario. After observing the results it can be easily stated that the people are ready for a change in system and will encourage the use of MIS in *excise department* for the payment of automobile tax. By using MIS for time saving and making the payment of automobile tax online the government will gain a great revenue as most people just don't pay their taxes because of the process that is been used right now.

## VII. Barriers involved in role of MIS in time saving techniques:

In the third part of the questionnaire the respondents were asked to identify the factors involved in limiting the role of management information system in government departments. The barriers involved are arranged as follows.

Large proportion (190 = 85.5%) rated that the lack of governments own interest towards the betterment of their departments in the field of MIS as one of most common and definite obstacle. The government is not yet taking MIS seriously in every aspect as they seem to be confused and divided in whether they should stay with the old approach or has the time arrived that the government should take prominent steps in the field of MIS.

In addition, (163 = 73.4%) blamed the end user as well in this case as they think that the common person doesn't think himself secure online with all his bank information is visible and he could easily be scammed by anyone. Lack of trust in the e-billing as they don't think of it as secure so they rather love to go with the flow and make MIS as their first priority.

Furthermore, (147 = 66.2%) respondents have seen technical and networking issues as well as lack of uniform standards as one of the major obstacles. Various studies have shown the negative relationships between perceived risk and constructs in the model. Dowling and Staelin (1994) indicated that the uncertainties, (that is perceived risk), may influence product adoption and evaluation (that is perceived usefulness). Moore and Benbasat (1991) have shown that a user interface that is complex may reduce the system's evaluation and adoption intention and it must be kept user friendly. Fu et al. (2006) shown that citizens may not receive a framework on the off chance that they see that the e-documenting framework is deficient in security highlights. Also, Featherman and Pavlou (2003) demonstrated that a framework that is dangerous and needs a ton of push to utilize might be seen as confronting execution issues and utilization vulnerabilities [11].

## VIII. Conclusion:

The need of MIS in time saving can be realized from its ambitions which are to develop an efficient system to maximize the effective and efficient use of data to management implementation [12]. Previous researchers have highlighted the importance of perceived hazard to the selection of e-filing. This research attempts to provide bits of knowledge into its facets, hence, giving useful contribution on the selection and evaluation of the e-filing system by users. It is predicted that a considerable lot of these risk facets will be huge. Among the dangers that could be critical are performance hazard, psychological hazard, time hazard and protection chance. Past studies have demonstrated that taxpayers tend to e-file near the expense deadline and this may lead to system crashes if the e-filing system isn't tailored to accommodate this trend. Psychological and time dangers could be prevalent for taxpayers who are not IT literate, they may get themselves frustrated or on edge if a considerable measure of time is spent learning about the e-filing system and after that find that the system does not work as they had hoped it would. Protection hazard may conceivably be a noteworthy hazard for e-filing selection this is because e-filing involves the transmission of taxpayer's confidential data through the Internet. The critical impacts of perceived usefulness, perceived convenience, and perceived credibility on behavioral goal were seen, no sweat of utilization applying a more grounded impact than either perceived usefulness or perceived credibility. The discoveries of this research give imperative strides to creating successful and proficient electronic taxpayer supported organizations when all is said in done and compelling electronic duty recording frameworks specifically. The examination affirmed that the MIS should be utilized all the more intensely in the choice procedure in the midst of emergencies. It was prescribed that the MIS units should be kept up to guarantee a free stream of information and sufficient utilization of MIS in choice generation. We could prescribe that appropriate introduction should be given to chiefs at all levels. Similarly, planning ventures must be sorted out to guarantee legitimate and sufficient utilization of MIS offices in creating and spreading information for better choices and efficient methods. However, attempting to make efficient and great level of arranging, coordination and control on the activities related to the usage of technology.